\documentclass[conference]{IEEEtran}
\IEEEoverridecommandlockouts

\usepackage{graphicx}
\usepackage{caption}
\usepackage{subcaption}
\usepackage{amsmath}
\usepackage{amssymb}
\usepackage{mathtools}
\usepackage{balance}
\usepackage{cite}
\usepackage{comment}
\usepackage{mathtools}
\usepackage{mathrsfs}

\newcommand{\qed}{\hfill\blacksquare}

\graphicspath{ {./Figures/} }
\DeclareGraphicsExtensions{.eps,.pdf,.png,.jpg,.gif,.jpeg,.pstex}
\allowdisplaybreaks

\newtheorem{lemma}{{\bf Lemma}}
\newtheorem{example}{{\bf Example}}
\newfont{\bb}{msbm10 scaled 1100}
\newcommand{\CC}{\mbox{\bb C}}

\newcommand{\dv}{{\bf d}}

\newcommand{\Wm}{{\bf W}}

\newcommand{\Zm}{{\bf Z}}

\newcommand{\Sc}{{\cal S}}

\newcommand{\muv}{\hbox{\boldmath$\mu$}}

\newcommand{\xsf}{{\sf x}}

\newcommand{\Ksf}{{\mathsf K}}

\newcommand{\Msf}{{\mathsf M}}
\newcommand{\Nsf}{{\mathsf N}}

\newcommand{\RED}{\color[rgb]{1.00,0.10,0.10}}

\usepackage{hyperref}
\hypersetup{
    bookmarks=true,         
    unicode=false,          
    pdftoolbar=true,        
    pdfmenubar=true,        
    pdffitwindow=false,     
    pdfstartview={FitH},    
    pdfnewwindow=true,      
    colorlinks=true,       
    linkcolor=red,          
    citecolor=green,        
    filecolor=blue,      
    urlcolor=blue           
}

\begin{document}
	
	\title{Cache-Aided Modulation for Heterogeneous Coded Caching over a Gaussian Broadcast Channel}
	
	\author{\IEEEauthorblockN{Mozhgan~Bayat\IEEEauthorrefmark{1}, Kai~Wan\IEEEauthorrefmark{1}, Mingyue Ji\IEEEauthorrefmark{2} and~Giuseppe~Caire\IEEEauthorrefmark{1}}
		\IEEEauthorblockA{ \IEEEauthorrefmark{1} Communications and Information Theory Group, Technische Universit\"{a}t Berlin, 10623 Berlin, Germany}
		\IEEEauthorblockA{ \{bayat, kai.wan, caire\}@tu-berlin.de}
		\IEEEauthorblockA{\IEEEauthorrefmark{2}University of Utah, Salt Lake City, UT 84112, USA, mingyue.ji@utah.edu}%
	}

	\maketitle
	
	\begin{abstract}
		Coded caching is an information theoretic scheme to reduce high peak hours traffic by partially prefetching files in the users local storage during low peak hours. 
		This paper considers  heterogeneous decentralized caching systems where users' caches and content library files 
		may have distinct sizes. The server  communicates with the users through a Gaussian broadcast channel.
		The main contribution of this paper is a novel modulation strategy to map the {\em multicast} messages 
		generated in the coded caching delivery phase  to the symbols of a signal constellation, 
		such that users can leverage their cached content to demodulate the desired symbols with higher reliability. 
		For the sake of simplicity, in this paper we focus only on ``uncoded'' modulation and symbol-by-symbol error probability. 
		However, our scheme in conjunction with multilevel coded modulation can be extended to channel coding over a larger block lengths.
	\end{abstract}

	
	\section{Introduction}
	\label{sec:introduction}
	Coded caching, originally proposed by Maddah-Ali and Niesen (MAN) in their seminal work \cite{maddah2014fundamental}, leads to an additional coded multicast gain compared to the conventional uncoded caching. In the MAN model, 
	a server has access to a library of $N$ files and is connected to $K$ users through an error-free shared link of unit capacity. 
	Each user is equipped with a cache of size equivalent to $M$ files. The MAN coded caching scheme consists of two phases: 
	placement and delivery.  During the placement phase, users partially store files from the library in their cache memories. 
	Of course, placement is agnostic of the future user demands. After the user demands are revealed, the server broadcasts a sequence of 
	multicast messages to the users. Such messages are computed as a function of the user demands, of the library files, and of the user cache content, 
	such that after receiving the multicast messages all users obtain their requested file with zero error probability (or vanishing error probability in the limit of 
	large file size). The placement can be done in a several manners distinguished by two characteristics, coded vs. uncoded and
	centralized vs. decentralized. Uncoded placement refers to the fact that segments of the library files are stored directly in the caches, and not
	more general functions thereof. It is known that for the MAN shared  link scenario uncoded placement is optimal within a factor of 2 \cite{exactrateuncoded}.
	Therefore, in this paper we consider uniquely uncoded placement.  A coded caching system is called {\em centralized} 
	\cite{maddah2014fundamental} if the server assigns the files segments to the users as a function of the number of users in the system. 
	In contrast, in the decentralized case \cite{decMAN}, each user individually and independently of the others fills up its cache 
	with bits from the library files without knowing how other users are in the system and which segment have been already cached by the other users. 
	A vast class of decentralized caching placements consists of {\em random independent caching}, where the set of library bits cached by each user $k$ be a random 
	variable $Z_k$  according  to some distribution $p_{Z_k}(\cdot)$ independent of the number of users $K$, and let the $\{Z_k : k = 1, \ldots, K\}$ be 
	statistically independent. 
	
	In a practice, it may be more realistic to consider the case where users and files have distinct  sizes (heterogeneous caching systems).
	In \cite{zeropadding}, the authors proposed a decentralized coded caching scheme with varying cache sizes by applying zero-padding to subfiles of different length to enable their encoding in a joint multicast message.
	Coded caching with distinct file sizes with uncoded placement was originally investigated in \cite{zhangfirst} where the 
	users could request a file multiple times. 
	They proposed  a caching scheme for different file sizes by considering random independent caching where the bits of each file are cached independently at random with a probability proportional to the file size.
	Further improvements on heterogeneous caching could be found in~\cite{zhangclosinggap,yenerd2d,OptimalDecFile}. 
	A common point of the existing heterogeneous caching schemes is that the delivery phases are based on clique-covering 
	method, which is a direct extension of the MAN delivery.
	\footnote{\label{foot:clique}Each transmitted multicast message in the delivery phase is a binary sum of a set of subfiles and useful to a subset of users, where each corresponding user knows all subfiles in the sum except one such that it can decode the remaining subfile.}
	
	In this paper,  we consider the implementation of a heterogeneous decentralized coded caching system 
	over a Gaussian broadcast channel, which is a more realistic model for the actual communication 
	physical layer than of the error-free capacitated shared link. Our main focus is to map the coded packets 
	generated in the caching delivery phase to a signal constellation. 
	\footnote{\label{foot:lattice index coding}This modulation with side information strategy was originally proposed in~\cite{latticeindexcoding} for index coding, where the authors~\cite{latticeindexcoding} considered how to modulate the index codes. The relationship between index coding and coded caching was discussed in~\cite{maddah2014fundamental,ontheoptimality}, and the main difference is that the stored content of each user is fixed in the index coding problem while the cache of each user can be designed in the caching problem (such that the ``worst'' cache configurations can be avoided.)} 
	In heterogeneous caching systems, the subfiles in each multicast message generated by a clique-covering method  
	may have distinct sizes, i.e., there is some inherent redundancy in each multicast message. 
	The idea is to leverage this redundancy in the modulation/demodulation step, such that the average symbol error rate of users can be reduced.
	Besides introducing the novel caching-modulation problem, our main contributions are
	\begin{itemize}
		\item We propose a novel modulation/demodulation strategy, where users can leverage their cached content to demodulate.
		\item We use that the {\it set partitioning} labelling proposed in~\cite{multilevel,ungerboeckpart1,ungerboeckpart2,Forney} is optimal (i.e., where the minimum distance is maximized) in our modulation with side information  strategy.
		\item We prove that the proposed cache-aided modulation scheme outperforms the conventional modulation scheme with zero padding.
	\end{itemize}

	\section{System model and problem setting}
	\label{sec:main}
	
	\subsection{System model}
	\label{sec:model}
	We consider a content delivery system with a server having access to a library of $\Nsf$ independent files $\Wm = \{W_1, W_2, \dots, W_{\Nsf}\}$ with distinct sizes.   For each $i\in [N]$, File $W_{i}$    has $F_i B$ bits where   $B = \sum_{j=1}^{\Nsf} |W_j|$ is the total library size in bits, 
	and $F_i = \frac{|W_i|}{B}$.  
	The server (e.g., a wireless base station) transmits a signal $x(t)$ to the users which receive $y_k(t) = \sqrt{\gamma_k} x(t)  + \nu_k(t)$, where $\nu_k(t)$ 
	is the  Additive White Gaussian Noise (AWGN) at the $k$-th receiver, with unit power, $x(t)$ is also normalized to have unit power, 
	and $\gamma_k$ denotes the receiver Signal-to-Noise Ratio (SNR). Without loss of generality we shall adopt the standard 
	complex baseband discrete-time model and since we focus on symbol-by-symbol demodulation we can omit the discrete time index and simply write
	$y_k = \sqrt{\gamma_k} x + \nu_k$ for a generic symbol at user $k$ receiver, use $X$ and $Y_k$ to denote the whole transmit and receive sequences over many symbols.
	Each user $k$ has a cache memory with size $\Msf_k$ bits where $\Msf_k \in [0, B]$. 
	We defined normalized cache sizes as $\mu_k = \Msf_k/B$. 
	The users have different cache sizes, without loss of generality, $\mu_1 \leq \mu_2 \leq \ldots \leq \mu_K$.
	
	The caching system comprises a placement and a delivery phase.
	In the placement phase users store contents from the library in a decentralized  manner  without any knowledge about demands. We define $\phi_k$ as the caching function for user $k$, which maps the library $\Wm$ to the cache content  $Z_k \overset{\Delta}{=} \phi_k(W_1,W_2,\dots, W_{\Nsf})$ for user $k$ with the content of all caches being denoted by $\Zm := (Z_1, Z_2, \dots, Z_{\Ksf})$.
	In the delivery phase, each user  requests one file from the library. We denote the file demanded by user $k$ as $d_k$ and demands of all users by $\dv  := (d_1,d_2,\dots,d_k)$.
	Given $ (\dv,\Zm)$, the server sends the codewords $X \in \mathscr{C}^L$, where $\mathscr{C}$ is a 
	$q$-dimensional signal constellation (i.e., a discrete set of points in $\CC^q$), and $L$ is the broadcast codeword length in terms of constellation symbols.\footnote{\label{foot:uncded}In this paper, for the sake of simplicity, we assume the modulation  is uncoded, i.e., 
		we let $q = 1$ and consider classical QAM/PSK signal constellations.}
	Upon receiving  $Y_k$, user $k$ needs to decode $W_{d_k}$ 
	from $Y_k$ and $Z_k$.
	
	Given the cache sizes of the users and the file sizes we design a shared link caching scheme 
	to fill the users' caches in the placement phase and to generate   broadcast messages $P$ of total size $R B$ bits 
	in the delivery phase. 
	The broadcast messages are designed such that each requested file $W_{d_k}$ 
	can be recovered from $(Z_k,P)$ for each $k \in [K]$. 
	This caching scheme is agnostic of the users' SNR and physical layer modulation. 
	The transmitter then maps the coded bits in $P$ into a sequence of $q$-dimensional constellation points, by dividing $P$ into labels of $m$ bits each, 
	and using these labels to index the $2^m$ points of the constellation $\mathscr{C}$, transmitted sequentially 
	over the Gaussian broadcast channel defined before.  
	The normalized load of the broadcast channel in terms of channel uses per $B$ bits 
	is given by $\frac{R}{m/q}$, where $R$ is the  load of the coded caching scheme as defined above and 
	$q/m$ is the spectral efficiency of the underlying physical layer modulation scheme, expressed in  
	bits per complex signal dimension. Such spectral efficiency depends on the physical layer modulation used, which in turns should be
	optimized with respect to the expected typical receiver SNR.\footnote{In general, the value of $m/q$ can be adapted depending on  
		the worst-case user SNR $\min \gamma_k$}
	
	The goal of the coded caching delivery scheme is to minimize $R$ while guaranteeing that each user demand can be satisfied, 
	subject to correct decoding of the multicast  messages.
	
	The objective of the physical layer modulation is to minimize the average symbol 
	error rate $\bar{T}$ among all users, where
	\begin{align}  
	\bar{T}= \frac{1}{K} \sum_{k\in [K]} \frac{S_k}{L_k},\label{eq:average symbol error rate}
	\end{align}
	where $S_k$ represents the number of symbols in $X$ which are useful to user $k$ 
	and decoded wrongly by user $k$, and  $L_k$ represents the number of symbols  in $X$ which are useful 
	to user $k$.

	\subsection{Decentralized  MAN caching scheme for heterogeneous network}  	\label{sec:man}
	
	In decentralized coded caching, during the  placement phase user $k$ independently fills his memory with $\mu_k F_iB$ bits of file $W_i$. 
	For each $\Sc \subseteq [\Ksf]$ and each $i\in [N]$, $W_{i,\Sc}$ represents the set of bits of $W_i$ which are uniquely cached by users in $\Sc$. Since $B$ is large enough, by the law of the large number, we have
	\begin{eqnarray}
	|W_{i, \Sc}|  = F_{i}B \left ( \prod_{j \in \Sc}  \mu_{j} \right ) \left ( \prod_{k \notin \Sc} (1-\mu_k)  \right ).
	\end{eqnarray}	
	In the delivery phase, for every non-empty subset of users $\Sc \subseteq [\Ksf]$, the server transmits  the following coded multicast messages
	\begin{equation}
	\label{eq:coded_multicast}
	P_{\Sc} = \bigoplus_{k \in \Sc} W_{d_k, \Sc\setminus\{k\}},
	\end{equation} 
	of length  $|P_{\Sc}| = \max_{k} |W_{d_k, \Sc \setminus k}|$, where enough zeros are added to the shorter subfiles to make their length to $ \max_{k} |W_{d_k, \Sc \setminus k}|$  in \cite{zeropadding}.  In addition, for all distinct demands $d_k$ the server must also broadcast
	directly to all users having requested file $d_k$ the subfiles $W_{d_k,\emptyset}$, which are requested but not cached at any user.
	Also, notice that the subfiles indexed by $\Sc = [K]$ is cached by everybody;  therefore, there is no need to transmit 
	subfiles $W_{d_k,[K]}$ to users.
	In this paper, for the simplicity of illustration, we use the decentralized MAN caching scheme. We will propose a novel cache-aided modulation scheme, which can be concatenated with any caching scheme based on clique-covering method.
	\section{Cached-aided modulation}
	
	The main idea of the proposed modulation scheme is that  the different lengths of the subfiles 
	in each multicast message $P_\Sc$ provide some side information to the users with larger cache size to demodulate $P_\Sc$. 
	We use a toy example to illustrate the idea.
	
	\begin{example}
		Consider a library of two files  $A = (101001010)$ of length  $9$ 
		bits and $B= (111001)$ of length $6$ bits. 
		Let $\Ksf = 2$ with cache memory $\Msf_1 = \Msf_2 = 5$ bits. 
		Two users randomly cache  $\mu_1 = \mu_2  = \frac{5}{15} = \frac{1}{3}$ of each file, and suppose that 
		the cache realizations is such that the subfile division is 
		$A = (A_{\emptyset}, A_{\{1\}},A_{\{2\}},A_{\{1,2\}} )=(101, 001, 010, \emptyset)$ and $B = (B_{\emptyset},B_{\{1\}}, B_{\{2\}},B_{\{1,2\}}) =( 11, 10, 01 ,\emptyset)$, i.e., the  cache contents are 
		$Z_1= \{ A_{\{1\}}, B_{\{1\}}, A_{\{1,2\}}, B_{\{1,2\}} \} = \{ 101, 10\}$ 
		and 
		$Z_2= \{ A_{\{2\}}, B_{\{2\}}, A_{\{1,2\}}, B_{\{1,2\}} \} = \{001, 01\}$.
		Let's assume the demands $d_1 = A$ and $d_2 = B$, and l  focus on the coded multicast message 
		\begin{align}
		P_{\{1,2\}} &= A_{\{2\}} \oplus B_{\{1\}} = 010 \oplus \xsf 10~ = 000,
		\end{align}
		where the symbol $\xsf$ denote a ``blank'' position due to the difference in length of the two subfiles. 
		Suppose that the 8PSK modulation constellation is used with labeling as shown in Fig.~\ref{fig:8PSK}, 
		such that $P_{\{1,2\}}$ is mapped directly onto the constellation point 
		indexed by the  3-bit label $P_{\{1,2\}}$.
		User $2$ has subfile $A_{\{2\}} = (010)$ in its cache memory 
		and wishes to decode subfiles $B_{\{1\}}= (10)$. Since 
		$|A_{\{2\}}| > |B_{\{1\}}|$, user $2$ knows the first bit in the label of the transmitted modulation symbol, which must be equal to 
		the first bit of $A_{\{2\}}$.
		Hence, it can demodulate the symbol by considering only the subconstellation of points whose first label bit is 0 (the blue points in 
		Fig.~\ref{fig:8PSK}). On the other hand, user $1$ does not know  any bit in the symbol label because its known subfile
		$B_{\{1\}}$ is shorter. Therefore, user $1$ must decode the symbol considering the whole constellation. 
		The minimum distance of the 8PSK constellation is  $2\sin{(\pi /16)}$ while the minimum distance of the ``blue'' subconstellation $2\sin{(\pi /4)}$. 
		It follows that user $2$ has a lower decoding error for the coded multicast message $P_{\{1,2\}}$ than user $1$.
		
		\begin{figure}[t]
			\centering
			\includegraphics[width=0.3\textwidth]{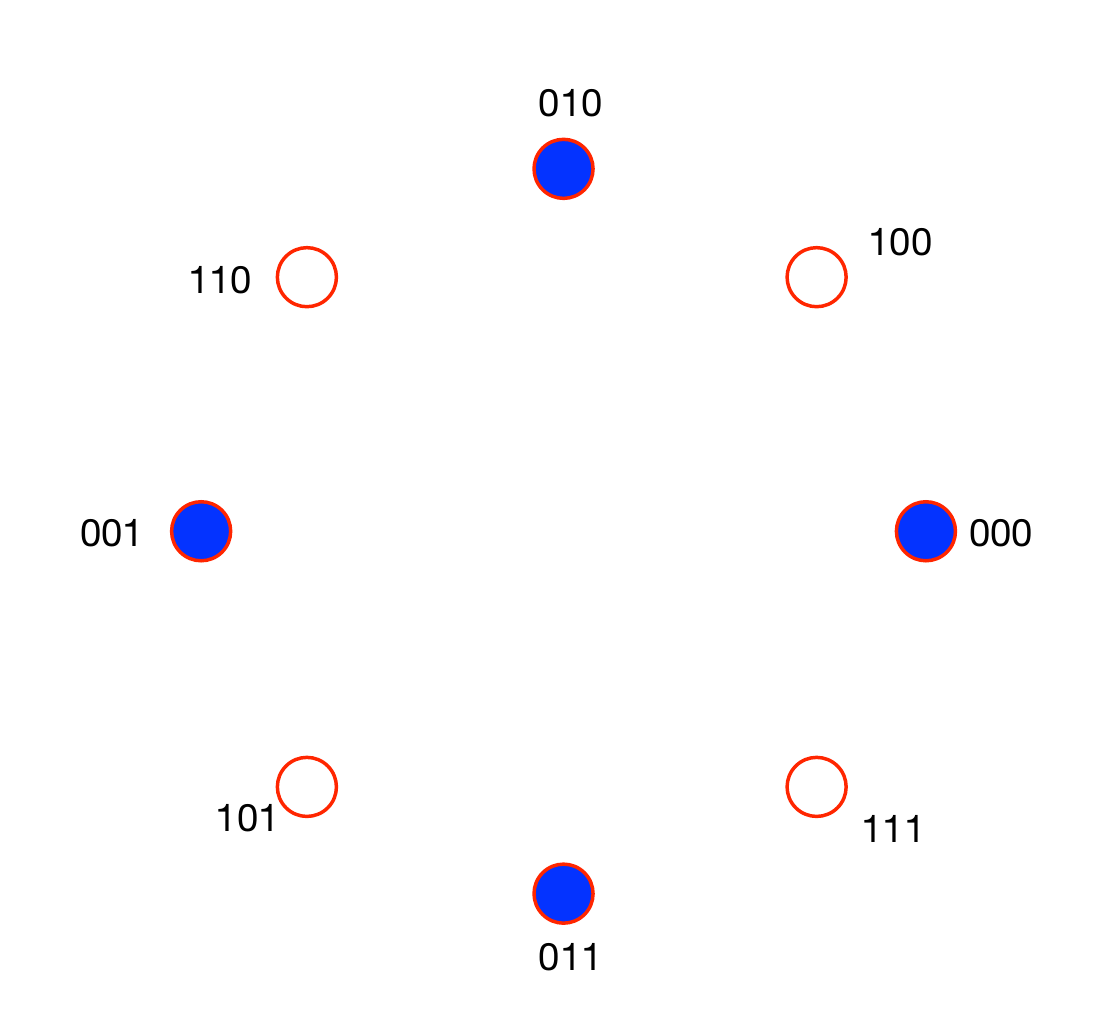}
			\caption{ Set partitioning labelling for  $8$-PSK }
			\label{fig:8PSK}
		\end{figure}

	\end{example}
	$\qed$
	
	
	In the following, we provide the general description of provide our proposed cache-aided modulation scheme. We define 
	$ \ell_{\Sc} = \max_{k \in \Sc} |W_{d_k, \Sc \setminus k}|$.
	To transmit requested subfiles, we 
	need to transmit  $n_{\Sc} = \frac{\ell_{\Sc}}{m} $ constellation symbols.\footnote{Since $B$ is large enough, we assume that $m$ divides $\ell_{\Sc}$.} First we divide each subfile $W_{d_k, \Sc \setminus k}$ into $n_{\Sc}$ pieces, each of which is denoted by   $W_{d_k, \Sc \setminus k}^i$ where $i \in [n_{\Sc}]$.
	\footnote{We also assume that $n_{\Sc}$ divides $|W_{d_k, \Sc \setminus k}|$. Because of this assumption, for user $k$ and $\Sc \in[K]$    we have $|W_{d_k, \Sc \setminus k}^i| =|W_{d_k, \Sc \setminus k}^j| $ for $\forall i,j \in [n_{\Sc}]$. }
	We generate one coded block 
	$$
	P_{\Sc}^i =  \bigoplus_{k \in \Sc} W_{d_k, \Sc\setminus\{k\}}^i, 
	$$  
	for all $i \in[n_{\Sc}]$ and then transmit  each block. Notice that each code block has size $m$, and therefore can be mapped directly onto
	a modulation point. We define $n_{\Sc,k}$ as the number of useful symbols to user $k$ among the symbols for $P_{\Sc}$.
	Notice that, in the proposed scheme, $n_{\Sc,k} = n_{\Sc}$ for all $k$. 	
	On the other hand, 
	the difference between the conventional zero padding scheme and our proposed scheme is  the way to partition each subfile into blocks.
	In the delivery phase of the zero padding scheme,
	we pad enough zero to at the end of $W_{d_k, \Sc \setminus k}$ to make it same length as the longest subfiles  $ \max_{k} |W_{d_k, \Sc \setminus k}|$ subfiles and divide it to $n_{\Sc}$ partitions denoted  as $W_{d_k, \Sc \setminus k}^i, i \in [n_{\Sc}]$. 
	Recall that $n_{\Sc,k}$ is the number of useful symbols among $n_{\Sc}$   symbols  for $P_{\Sc}$. Notice that,  in the zero padding scheme, $n_{\Sc,k}$ might be different for users in $\Sc$ and in general  we have $n_{\Sc,k} \leq n_{\Sc}$.
	
	\begin{figure}
		\centering
		\includegraphics[width=0.35 \textwidth]{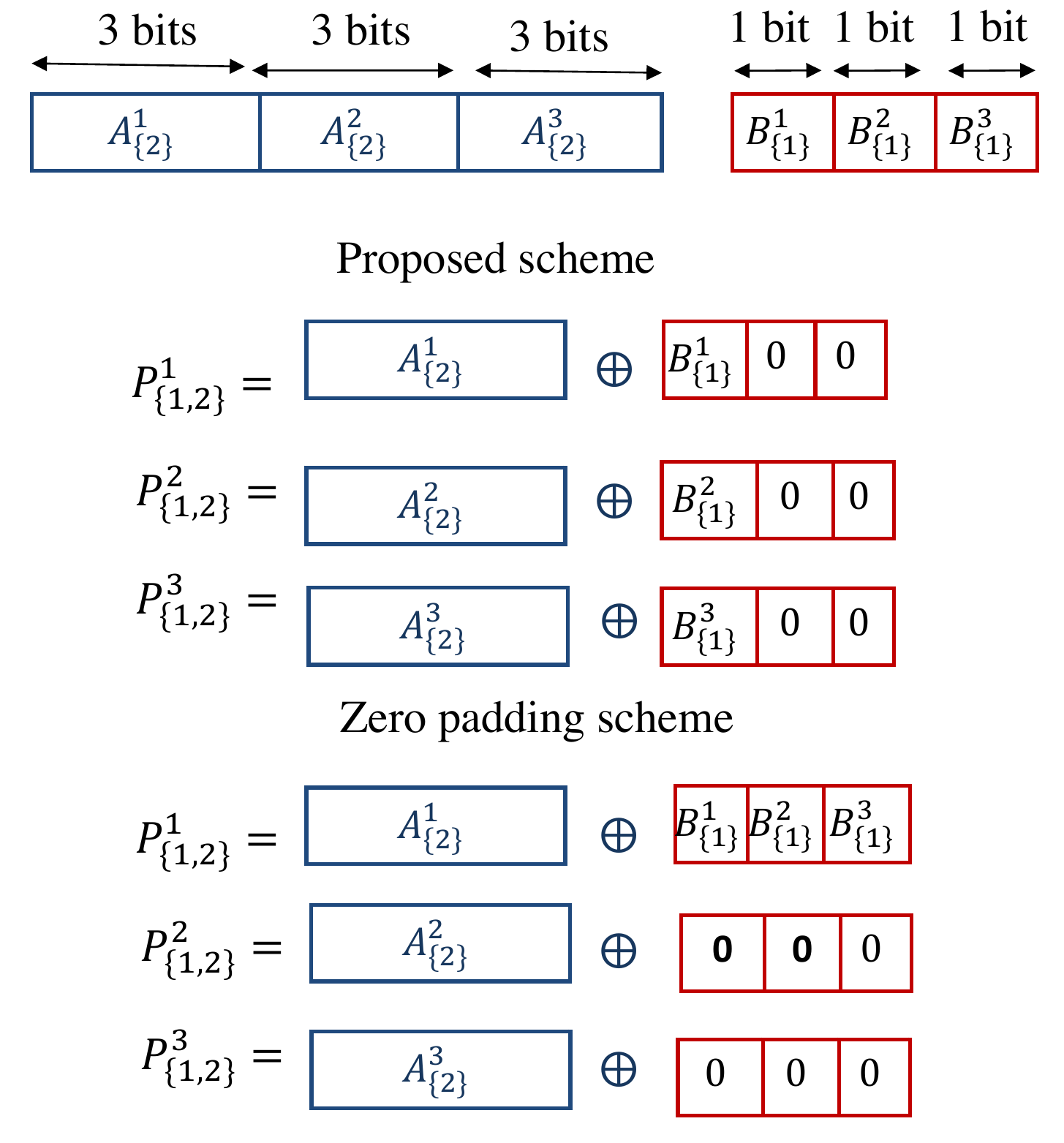}
		\caption{proposed  and zero padding scheme}
		\label{fig:comparsion}
	\end{figure}
	In Fig. \ref{fig:comparsion}, comparison between zero padding scheme and proposed scheme is illustrated through an example. In coded multicast message we would like to transmit subfiles $A_{\{2\}}$ of length $9$ bits and subfiles $B_{\{1\}}$ of length $3$ bits. Consider $m = 3$, for our proposed scheme we first divide $A_{\{2\}}$ and $B_{\{1\}}$ into 3 blocks of equal size and then pad with zeros to create blocks of length $3$.  After this per symbol padding, we encode them together as messages $P_{\{1,2\}}^1$ and $P_{\{1,2\}}^2$ and $P_{\{1,2\}}^3$. 
	In conventional zero padding, we pad with zeros the whole subfile $B_{\{1\}}$ to create a subfile of size $9$ bits. 
	Then we divide both subfiles into $3$ blocks and encode each block as messages $P_{\{1,2\}}^1$  and $P_{\{1,2\}}^2$ and $P_{\{1,2\}}^3$.
	In one word, the proposed scheme uses a zero padding on a ``symbol level", while the conventional scheme uses a zero padding on a ``subfile level".
	\subsection{Derivation of the uncoded symbol error rate}
	The  error probability for $2^{m}$-PSK  is bounded as follows \cite{proakis2001digital}
	\begin{equation} \label{eq:error-psk}
	P_{\rm e, 2^{m}-{\rm PSK}}  \leq 2Q \left ({\frac{d_{\rm min}}{\sqrt{2N_0}}}\right),
	\end{equation}
	where $d_{\rm min}$ is the minimum distance between any two data symbols in a signal constellation and the Q-function is defined as
	\begin{equation}
	Q(x) = \frac{1}{\sqrt{2 \pi}} \int_{x}^{\infty} e^{-\frac{u^2}{2}} d u.
	\end{equation}

	Since with the proposed binary labeling some users will have some of the most significant bits (leftmost in the label arranged from left to right) known, as seen in the example, we wish to use a binary labeling such that the $d_{\rm min}$  of the sub constellation indexed by the label with fixed first $n$ most significant bits is maximized for any configuration of the $n$ bits. This is known to be the labeling by set partitioning, well-known in the coded modulation literature  \cite{multilevel,ungerboeckpart1,ungerboeckpart2, Forney}. For this labeling for the $2^m$-PSK modulation we have 
	
	\begin{equation}
	d_{{\rm min},n} = 2 \sin \left (\frac{ \pi}{2^{m-n}} \right )d.  \label{eq:psk}
	\end{equation}
	
	A similar reasoning applied to QAM constelaltions of size $2^m$ obtained by carving $2^m$ from the infinite squared grid on the complex plane. In this case we have 
	
	\begin{equation}
	P_{\rm e, 2^{m}-{\rm QAM}}  \leq \left(  1- \left[ 1- 2Q \left({\frac{d_{\rm min}}{\sqrt{2N_0}}}\right) \right]\right)^2 
	\leq 4Q\left ({\frac{d_{\rm min}}{\sqrt{2N_0}}}\right).
	\end{equation}
	
	and the labeling by set partitioning yields the subconstelaltion minimum distance
	
	\begin{equation}
	d_{{\rm min}, n} = (\sqrt{2})^n d.  \label{eq:qpsk}
	\end{equation}

\begin{figure}[t]
		\centering
		\includegraphics[width=0.3\textwidth]{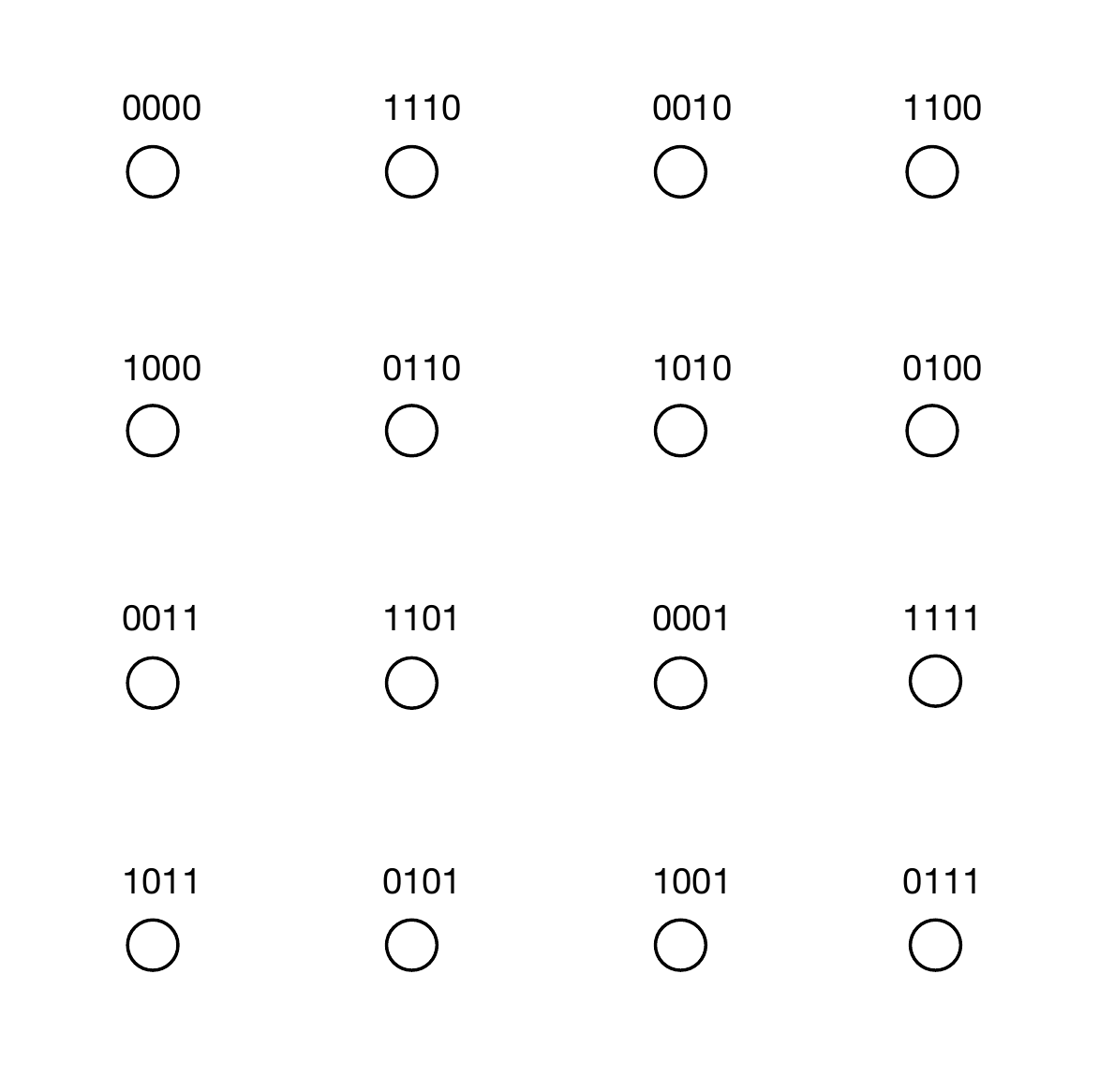}
		\caption{ Set partitioning labelling for $16$-QAM}
		\label{fig:qamproof}
	\end{figure}

	Fig.  \ref{fig:qamproof}  shows a typical labeling by set partitioning for the $16$-QAM constellation. 
	One can check that for any number of $n = 1,2,3$ known most significant bits the minimum distance of the resulting sub constellation satisfies \eqref{eq:qpsk}.

	The next step is to derive the error probability of decoding subfile   $W_{d_k, \Sc\setminus\{k\}}^i$, in coded multicast $P_\Sc^i$,
	at receiver user $k$. In order to do that, we first need to calculate how many bits are known to each receiver.
	The number of  bits are known to user $k$ in the multicast message $P_{\Sc}^i $  is equal to $ \max_{ k\in \Sc}|W_{d_{k}, \Sc \setminus k}^i| - |W_{d_{k}, \Sc \setminus {k}}^i|$ 
	since user $k$ has all other subfiles in his cache memory.
	By using \eqref{eq:error-psk} and \eqref{eq:psk}, the error probability of decoding $W_{d_k, \Sc\setminus\{k\}}^i$ with modulation $2^{m}$-PSK is given by
	\begin{align}
	\label{eq:prob_error}
	& P_{e,k, \Sc } =   \\
	& 2Q\left(  \sqrt{2 \gamma_k \sin^2 \left(\pi / 2^{m -\max_{j \in \Sc} |W_{d_{j}, \Sc \setminus {j}}^{i_0}| + |W_{d_{k}, \Sc \setminus {k}}^{i_0}| } \right) }\right ),   \nonumber
	\end{align}
	where $\gamma_k$ denotes the SNR at receiver of  user $k$ and $i_0 =1$. Notice  that $|W_{d_{k}, \Sc \setminus {k}}^i|$ is invariant regarding to index $i$. In other words,  $|W_{d_{k}, \Sc \setminus {k}}^i| = |W_{d_{k}, \Sc \setminus {k}}^{i_0}|$ where $\forall i \in[n_{\Sc}] $. 
	After calculating $P_{e,k, S} $ for $\forall i \in [n_{\Sc, k}]$
	The total number of  useful symbols for user $k$ is given by 
	\begin{equation}
	L_k = \sum_{\Sc : k \in \Sc} n_{\Sc, k}.
	\end{equation}
	The average total number of symbol errors among the symbols useful to user $k$ is give by
	\begin{eqnarray}
	S_k = \sum_{\Sc : k \in \Sc} n_{\Sc,k}  P_{e,k,\Sc}.
	\end{eqnarray}
	For each user $k$  average symbol error rate is defined as 
	\begin{equation}  
	\bar{T}_k=    \frac{S_k}{L_k}. \label{eq:ser}
	\end{equation}
	
	Finally,   $\bar{T}$ in  (\ref{eq:average symbol error rate}) is obtained by using  above equations.
	\begin{lemma} \label{lemma:psk}
		The proposed  cache-aided modulation achieves lower or equal symbol error rate $\bar{T_k}$ \eqref{eq:ser}  for all user $k$ than 	conventional zero-padding scheme.
		\begin{IEEEproof} By considering file sizes are very large, first we derive $\bar{T}_k$ for user $k$ for zero padding scheme. In zero padding scheme, for any $\Sc$ and any user $k \in \Sc$ we have two cases: first case if $i\in[n_{\Sc, k}]$ we have $|W_{d_k,\Sc\setminus k}^i |=m$ and second case if $ i \in [n_{\Sc}] \setminus [n_{\Sc, k}]$ we have    $|W_{d_k,\Sc\setminus k}^i |=0$. In any useful symbol for user $k$ with indexes in $[n_{\Sc,k}]$  the number of knows bits for user $k$ is zero,
			i.e. $\max_{j \in \Sc} |W_{d_{j}, \Sc \setminus {j}}^{i_0}| - |W_{d_{k}, \Sc \setminus {k}}^{i_0}| =0$. In zero padding scheme,  for all user $k$ for all $\Sc \subseteq [K]$ we have $P_{e,k } = P_{e,k,\Sc} $,  where $P_{e,k} = 2Q\left(  \sqrt{2 \gamma_k \sin^2 \left(\pi  \right) }\right)$. This implies $\bar{T}_k ^{\rm zp}= P_{e,k}$, where $\bar{T}_k ^{\rm zp}$ is denoted the uncoded symbol error rate in  \eqref{eq:ser}  for user $k$.
			 $\bar{T}_k^{\rm p}$ is denoted the symbol error rate for user $k$ for proposed scheme is given by
			\begin{subequations}
				
				\begin{align}
				\bar{T}_k^{\rm p} &= \frac{ \sum_{\Sc : k \in \Sc} n_{\Sc, k}P_{e,k,\Sc} }{ \sum_{\Sc : k \in \Sc} n_{\Sc, k}} \label{eq:const1} \\
				& \leq  \max_{\Sc} P_{e,k,\Sc} \label{eq:const2} \\
				& \leq  2Q\left(  \sqrt{2 \gamma_k \sin^2 \left(\pi   \right) }\right) \label{eq:const3} \\
				& = \bar{T}_k ^{\rm zp}.
				\end{align}
			\end{subequations}
			
			For any received signal in any receiver for proposed scheme the number of known bits and $d_{\rm min}$  areat least as big as the number of  known bits for zero padding ones.
		\end{IEEEproof} 		
	\end{lemma}
	
	\section{Simulation results}
\begin{figure}[t]
		
		\centering
		\includegraphics[width=0.5\textwidth]{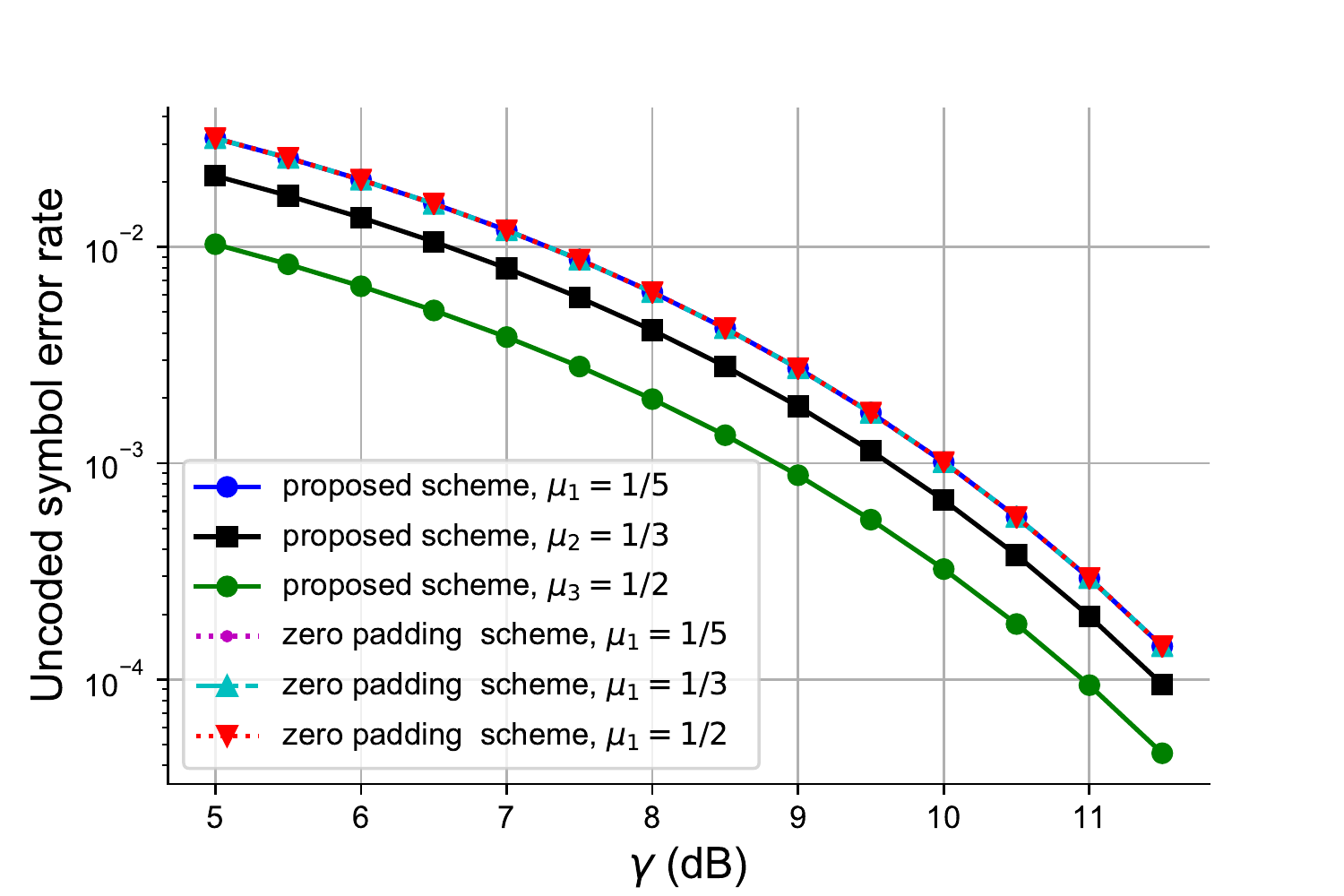}
		\caption{$\bar{T}_k$ symbol error rate  of each user  versus SNR for  a network with parameters   $K=3$ , $\muv = (1/5,1/3,1/2)$ and for each user $\gamma_k = \gamma$. }
		\label{fig:syb_rate}
	\end{figure}

\begin{figure}[t]
		
		\centering
		\includegraphics[width=0.45\textwidth]{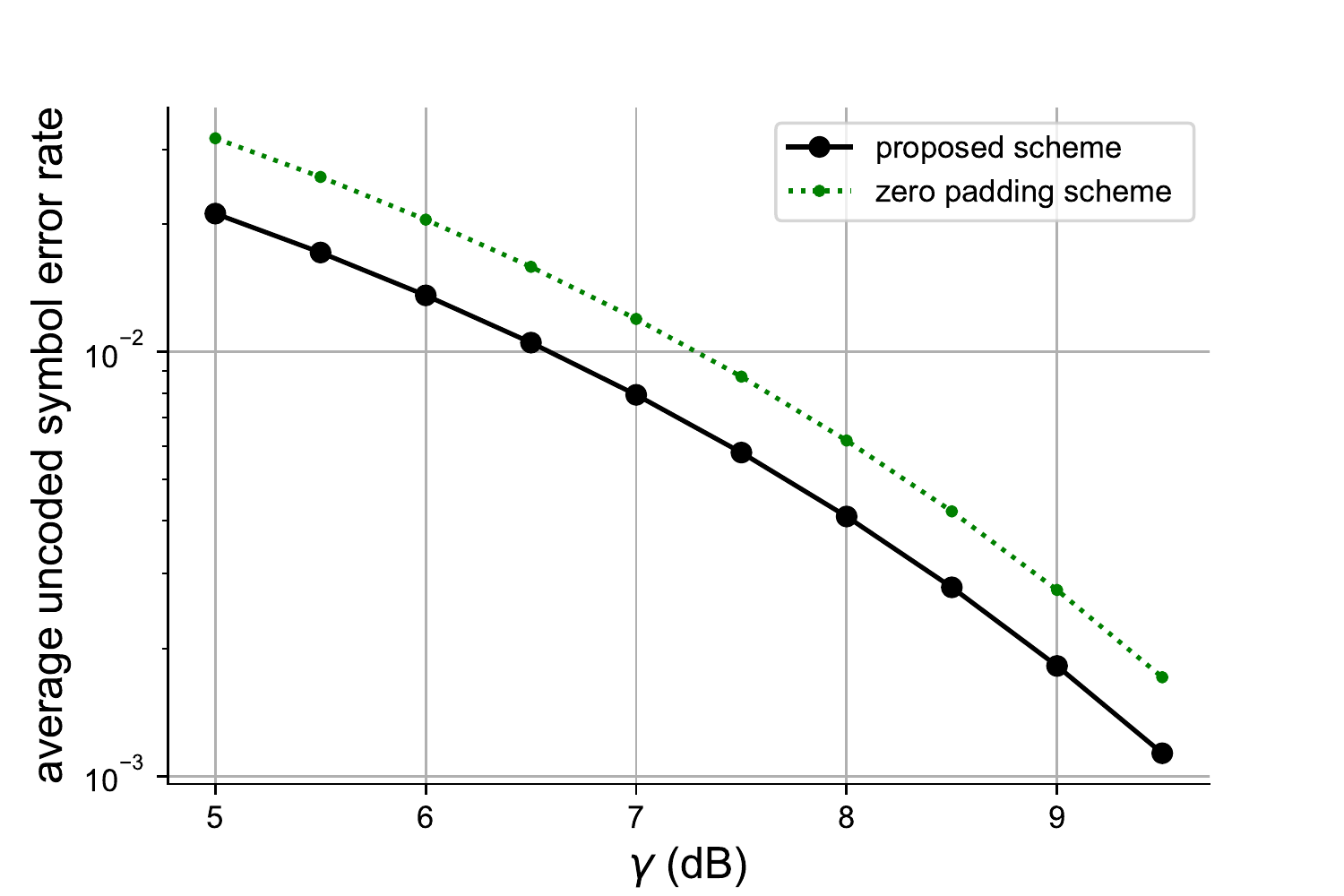}
		\caption{$\bar{T} $ average  symbol error rate   versus SNR for  a network with parameters   $K=3$ , $\muv = (1/5,1/3,1/2)$ and for each user $\gamma_k = \gamma$. }
		\label{fig_avg}
	\end{figure}

	In Fig.  \ref{fig:syb_rate} and Fig. \ref{fig_avg} ,  $\bar{T}_k$ for each user $k$ in \eqref{eq:ser}  and $\bar{T}$ in \eqref{eq:average symbol error rate} versus SNR are plotted. 
	In Fig.  \ref{fig:syb_rate} we compare $\bar{T}_k$ in \eqref{eq:ser} 
	for different users for proposed and zero padding scheme. 
	User $1$ has lower cache size among other users, which implies  that for this user our scheme does not have any improvement compare to zero padding scheme, the symbol error rate of user $1$  with $\mu_1 =1/5$ for both case are same.
	User $2$ has larger cache size   $\mu_2=1/3$,  symbol error rate  of   proposed scheme improves compare to zero padding one. Our proposed algorithm achieves noticeable gain for user $3$  with  cache size $\mu_3=1/2$ who has biggest cache size among other users. 
	
	\section{conclusion}
	
	In this paper, for heterogeneous caching systems, we  proposed a novel scheme to map the multicast messages generated by the clique-cover delivery method onto physical layer modulation symbols such that users with larger subfiles can take advantage of the known bits in the constellation
	labels, effectively restricting their detection problem to a sub-constellation of increased minimum distance. 
	We showed that our scheme achieves better or equal symbol error rate for all users with respect to the 
	conventional zero-padding scheme, which does not use cache-aided side information for the demodulation process to any user.
	In addition, it can be  seen that the best labeling for our scheme is the well-known binary set partitioning labeling, widely used in 
	standard coded modulation techniques. Finally, it is possible to extend the proposed scheme to constellation constructions of longer dimension $q$ using the technique
	of multilevel coded modulation \cite{Huber}, and replace uncoded error rate with coded block error rate using finite-length coding results \cite{polyanskiy}.
	In this paper we have focused on the uncoded case for simplicity and for the sake of space limitation, 
	while the full characterization of the achievable tradeoff between coding rate and block error rate is work in progress.


	
	\balance 
	{
		\bibliographystyle{IEEEtran}
		\bibliography{references}
	}

\end{document}